# Optical design of the SOXS spectrograph for ESO NTT

Ricardo Zanmar Sanchez*[a], Matteo Munari[a], Adam Rubin[h], Sagi Ben-Ami[b], Anna Brucalassi[f], Hanindyo Kuncarayakti[j,k], Jani Achrén[l], Sergio Campana[c], Riccardo Claudi[d], Pietro Schipani[e], Matteo Aliverti[c], Andrea Baruffolo[d], Federico Biondi[d], Giulio Capasso[e], Rosario Cosentino[g,a], Francesco D'Alessio[i], Paolo D'Avanzo[c], Salvatore Scuderi[a], Fabrizio Vitali[i], José Antonio Araiza-Duran[m], Iair Arcavi[n], Andrea Bianco[c], Enrico Cappellaro[d], Mirko Colapietro[e], Massimo Della Valle[e], Oz Diner[h], Sergio D'Orsi[e], Daniela Fantinel[d], Johan Fynbo[o], Avishay Gal-Yam[h], Matteo Genoni[c], Ofir Hershko[h], Mika Hirvonen[p], Jari Kotilainen[j], Tarun Kumar[k], Marco Landoni[c], Jussi Lehti[p], Gianluca Li Causi[q], Luca Marafatto[d], Seppo Mattila[k], Giorgio Pariani[c], Giuliano Pignata[m], Michael Rappaport[h], Davide Ricci[d], Marco Riva[c], Bernardo Salasnich[d], Stephen Smartt[r], and Massimo Turatto[d]

[a]INAF - Osservatorio Astrofisico di Catania, Catania, Italy, [b]Harvard-Smithsonian Center for Astrophysics, Cambridge, USA, [c]INAF - Osservatorio Astronomico di Brera, Merate, Italy, [d]INAF - Osservatorio Astronomico di Padova, Padua, Italy, [e]INAF - Osservatorio Astronomico di Capodimonte, Naples, Italy, [f]European Southern Observatory, Garching, Germany, [g]INAF - Fundación Galileo Galilei, Breña Baja, Spain, [h]Weizmann Institute of Science, Rehovot, Israel, [i]INAF - Osservatorio Astronomico di Roma, Rome, Italy, [j]FINCA - Finnish Centre for Astronomy with ESO, Turku, Finland, [k]University of Turku, Turku, Finland, [l]Incident Angle Oy, Turku, Finland, [m]Universidad Andres Bello, Santiago, Chile, [n]Tel Aviv University, Tel Aviv, Israel, [o]Dark Cosmology Centre, Copenhagen, Denmark, [p]ASRO - Aboa Space Research Oy, Turku, Finland, [q]INAF - Istituto di Astrofisica e Planetologia Spaziali, Rome, Italy, [r]Queen's University Belfast, Belfast, UK

**ABSTRACT**

An overview of the optical design for the SOXS spectrograph is presented. SOXS (Son Of X-Shooter) is the new wide-band, medium resolution (R>4500) spectrograph for the ESO 3.58m NTT telescope expected to start observations in 2021 at La Silla. The spectroscopic capabilities of SOXS are assured by two different arms. The UV-VIS (350-850 nm) arm is based on a novel concept that adopts the use of 4 ion-etched high efficiency transmission gratings. The NIR (800-2000 nm) arm adopts the '4C' design (Collimator Correction of Camera Chromatism) successfully applied in X-Shooter. Other optical sub-systems are the imaging Acquisition Camera, the Calibration Unit and a pre-slit Common Path. We describe the optical design of the five sub-systems and report their performance in terms of spectral format, throughput and optical quality. This work is part of a series of contributions[1-9] describing the SOXS design and properties as it is about to face the Final Design Review.

## 1. INTRODUCTION

SOXS (Son of X-Shooter) will be the new medium resolution spectrograph for the NTT telescope at la Silla, expected to start operations in 2021. SOXS was designed to cover the wavelength range from 350 to 2000 nm using two different spectrographs in the UV-VIS (350-850 nm) and NIR (800-2000 nm). Initially, as the name implies, it was conceived to be entirely based on the successful X-Shooter design implemented at the VLT telescope, however, the UV-VIS arm has evolved to a novel approach based on custom-made ion-etched gratings[7].

*zanmar@oact.inaf.it; phone +39 095 733-2206

The five subsystems described here are related as follows: incoming light from the telescope is split and relayed to the NIR and UV-VIS spectrographs by the Common Path. To perform successful observations, SOXS also includes a Calibration Unit that removes the instrumental signature from the spectra with light sources of known flux and wavelength. SOXS will also produce science grade photometry with the Acquisition Camera.

The most important optical requirements satisfied by the current design are:

- resolution better than 3500 for a 1-arcsec slit
- 80% enclosed energy distribution better than 0.4 arcsec for the whole wavelength range
- Overall throughput from telescope to detector better than 20% for both NIR and UV-VIS arm

In the following sections, we describe five subsystems involved in the optics of SOXS: Common Path (CP), UV-VIS spectrograph, NIR spectrograph, Calibration Unit (CU) and Acquisition Camera (AC).

## 2. COMMON PATH

The first optical component of the CP is a dichroic (Figure 1) that splits the spectral band into the 350-850 nm and 850-2000 nm wavelength ranges that feed the UV-VIS and NIR spectrographs, respectively. The dichroic is used in reflection for the UV-VIS light to allow for better reflection coatings. It is used at an angle of 15 degrees and has a small wedge to reflect out of the slit its ghost image.

Both NIR and UV-VIS arms of the CP have the function of re-imaging the light from the telescope focal point into their respective spectrographs slits while changing the beam F/# from 11 to a fast 6.5 in order to allow for compact spectrographs. Both arms share a similar concept, with two folding mirrors and doublets to change the F/#. The last optical element in both arms is a field lens that re-positions the exit pupil to match that of the telescope. One of the two mirrors in each arm is allowed to tip and tilt in order to compensate for mechanical flexures as the telescope moves.

The entrance window of the NIR spectrograph (see Section 4) is optically part of the CP. It is made out of Silica and allows light to enter the NIR spectrograph dewar. The window is tilted to reflect out of the slit the ghost image produced by this element. The NIR arm doublet is mounted on a slide motor to allow for an independent focusing mechanism in addition to the secondary mirror of the telescope.

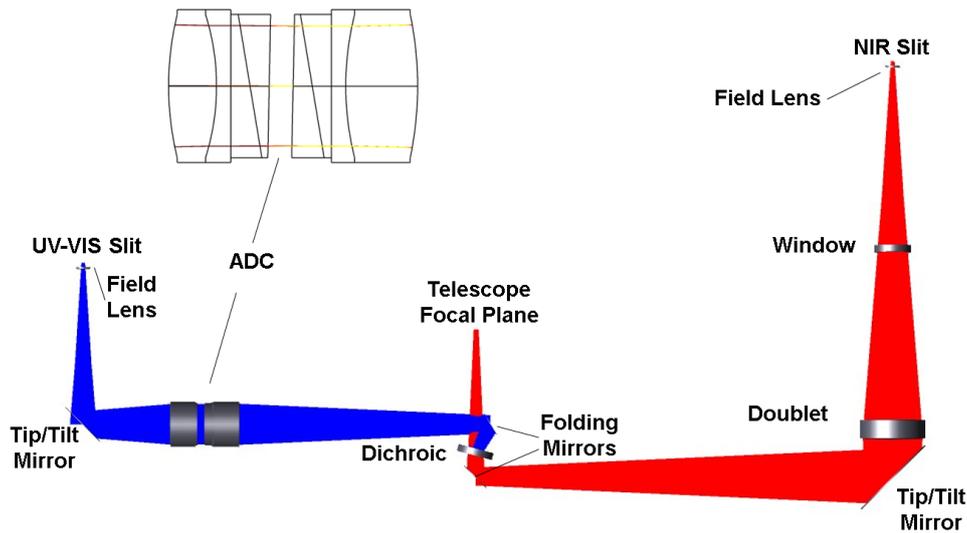

Figure 1 Common Path layout. In Blue, the UV-VIS arm and in Red the NIR arm. Upper left inset is an enlargement of the ADC.

The UV-VIS arm of the CP includes an atmospheric dispersion corrector (ADC) that is permanently located in a collimated beam produced by two doublets. The ADC is composed of two identical pair of prisms made out of BAL15Y and S-FPL51Y cemented to the doublets, see Figure 1, upper left inset. Each pair is allowed to counter-rotate to adjust the correction according to the observing zenith angle. Figure 2 are the resulting ADC angles as zenith angle is varied under different atmospheric conditions (pressure P, temperature T). When ADC angle is 90°, prisms are anti-aligned and the dispersion of one cancels the other. For ADC angle 0°, prisms are aligned for maximum dispersion. A model of the form

$$\text{prism\_angle} = \text{acos}(\ (3.6e-1 - 1.2e-3 \times T + 4.4e-4 \times P) \times \tan(\text{zenith\_angle})\ ) \quad (1)$$

has been fitted (plotted as solid lines) in Figure 2 and will be implemented in the software. The atmospheric dispersion in the NIR wavelength range is expected to be less than 0.5 arcsec and it was regarded as acceptable, therefore no ADC is included in this arm. A summary of the CP main characteristics is in Table 1.

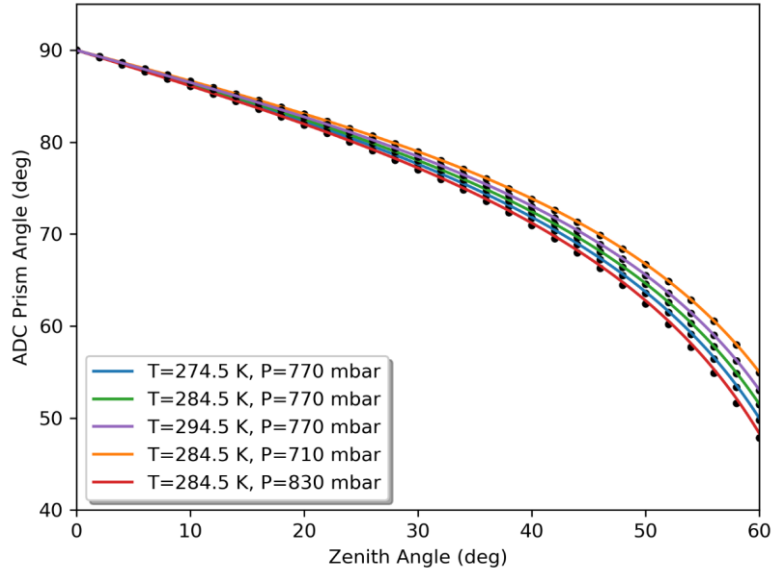

Figure 2. ADC prism angle. The points are prism angles found by optimization in ZEMAX. The solid lines are best fitting models.

The image quality throughout the slit, as measured by the spot size, is less than 10µm. When accounting for possible manufacturing and positioning errors in Monte Carlo simulations, the spot size is still under 20µm (a $5^{th}$ of an arcsec). This holds for both of the CP arms.

The expected throughput of the CP in both arms is ~90% assuming efficiencies for reflective coatings and antireflection coatings of 99%. Dichroic efficiency was simulated by vendors.

Table 1. Common Path main characteristics.

|  | **UV-VIS** | **NIR** |
|---|---|---|
| Wavelength Range | 350-880 nm | 800-2000 nm |
| Input F/N | F/11 (telescope) | F/11 (telescope) |
| Output F/N | F/6.5 | F/6.5 |
| Field of View | 12x12 arcsec | 12x12 arcec |
| Scale | 110 µm/arcsec | 110 µm/arcsec |
| ADC | Yes | No |

We have checked the image quality and position of the CP as the temperature departs by +/-10C from the nominal 10C. Materials and kinematic mounts[2] were chosen in order to preserve the image position on the slits. Small defocusing is expected of up to 20µm due to temperature variations but it can be adjusted by moving the telescope secondary mirror. Any relative defocus between the NIR and UV-VIS arm can be compensated by moving the NIR arm doublet.

## 3. UV-VIS SPECTROGRAPH

The spectrograph MITS[7] (Multi-Imaging Transient Spectrograph) dedicated to the UV-VIS wavelengths is a novel concept inspired in the MOONS[12] spectrograph. It uses dichroics to split incoming light, transmission gratings as the dispersers and above all, the so called Wonder-Camera[12] that consists of only 2 lenses and a mirror.

Figure 3 is the optical layout for MITS. The incoming beam from the slit is collimated with an off-axis section of a parabolic mirror and then split with three dichroics as follows: the first dichroic splits the light into u and g, in transmission, r and i in reflection, and then each subsequent beam is further splitted to create the four quasi-orders u, g, r, i. Each of the four beams finds its corresponding disperser: a custom made ion-etched grating. The dispersed beams are re-imaged onto the detector (Figure 4, left) by the camera, which consists of an aspheric meniscus corrector, an aspheric mirror and an aspheric flat field lens. The latter is made of Silica and is the window of the CCD cryogenic chamber.

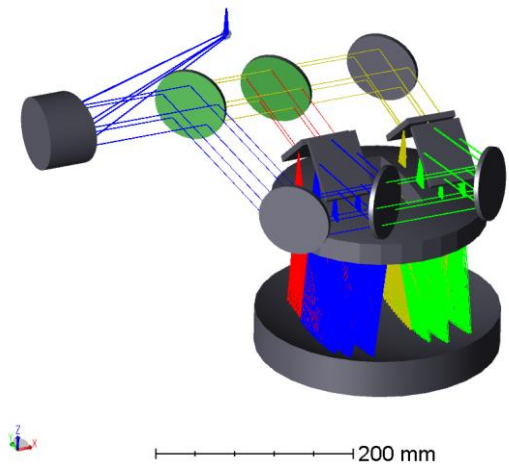

Figure 3. UV-VIS spectrograph optical layout

The adopted ion-etched gratings have an efficiency of ~85% in all four channels. The expected throughput from sky to detector is ~30% assuming efficiencies for narrow and broadband reflective coatings of 98.5% and 99% respectively, and 99% efficiency for AR coatings. Grating and dichroic efficiencies were simulated by vendors.

The image quality as measured by the 80% enclosed energy is better than 0.4 arcsec in both the spatial and spectral directions and satisfies the requirement (see Figure 4, right).

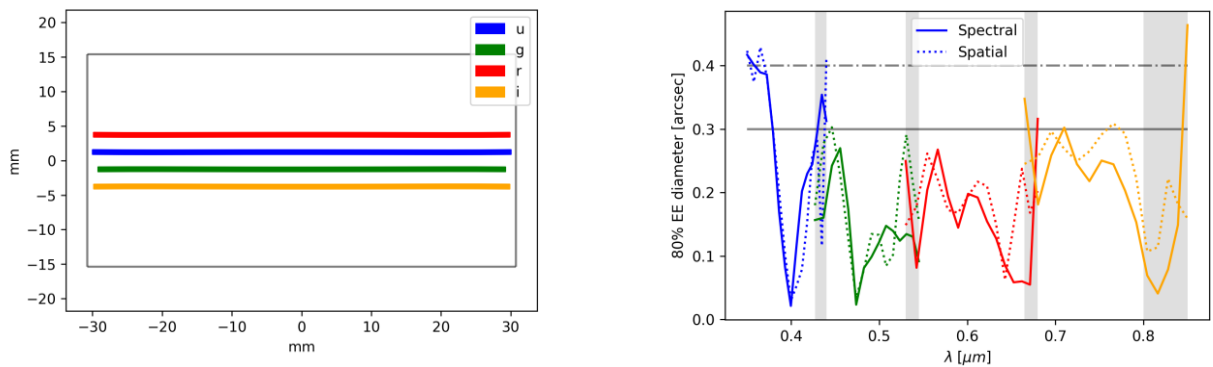

Figure 4. Left, spectral format on the detector. Right, enslitted energy diameter.

There are mainly four potentially problematic ghost images, three involve the field flattener lens and one the meniscus corrector. Fortunately, all but one do not overlap with the spectra. Furthermore, the overlapping ghost, formed by the reflection on the detector and front of the field flattener, is defocused and its relative intensity estimated to be only $10^{-6}$ and therefore negligible.

## 4. THE NIR SPECTROGRAPH

This spectrograph, as in X-Shooter, is based on the 4C concept which is a variation of the white pupil spectrograph where the dispersed beam overlaps again to form a second pupil (hence 'white'). 4C stands for Collimator Compensation of Camera Chromatism, where in our case, the chromatism produced by the camera is corrected by the meniscus of the collimator. The two main advantages of this approach are 1) the high efficiency due to the grating operating close to Littrow (4 deg off-plane in our case) and 2) compactness of the spectrograph because both, the crossdisperser and camera, can be placed near pupils.

The optical layout is presented in Figure 5. After the Common Path refocuses the object on the slit, the beam is folded by a mirror to enter the collimator, an off-axis Maksutov telescope. The corrector lens of the collimator introduces negative chromatism to counteract that introduced later by the camera. The collimated beam of size ~50mm enters three Cleartran prisms used as crossdispersers before reaching the reflecting grating. After the second pass through the prisms and collimator, the dispersed beam is focused by the collimator mirror onto the folding mirror which has power and forms the second pupil. The camera is placed near this pupil allowing for it to be compact. It contains three singlets, one aspherical surface. To minimize background noise, the NIR spectrograph is contained in a vacuum vessel with a working temperature of 150K. An Infrasil filter in the camera prevents light above 2000 nm from reaching the detector.

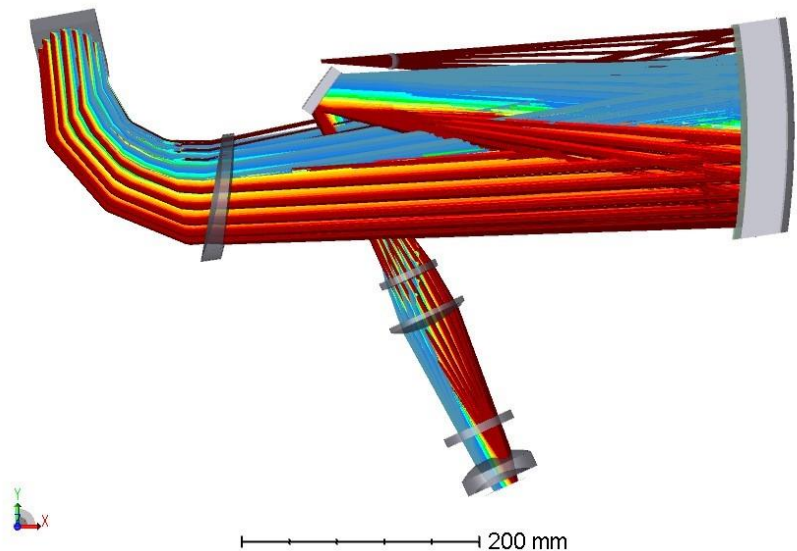

Figure 5. NIR spectrograph optical layout.

This kind of spectrograph does not suffer from vignetting and thus greatly compensates for losses in the double pass. The expected throughput is ~26% from sky to detector assuming a total filter transmissivity of 95% and grating efficiency from a possible manufacturer.

The three Cleartran prisms with apex angles 20˚, 20˚ and 26˚ provide an inter-order gap separation from 10 to 80 pixels (see Figure 6, left). The base of the prism never exceeds 50mm and should be feasible to manufacture.

The optical quality, expressed as the 80% enslitted energy, is presented in Figure 6, right. The detector is slightly tilted and can observe the spectrum from 800 to 2000nm in 15 orders (with a small gap between 1833 and 1842 nm), Figure 6, left. The first pass through the prisms before the grating introduces spectral line tilt of as much as 14° but the reduction pipeline will be able to correct for it.

The average spectral resolution is ~5300 for a 1 arcsec slit. It was estimated by using the geometrical size of the slit on the detector, taking into account the diameter of the enslitted energy, for each wavelength.

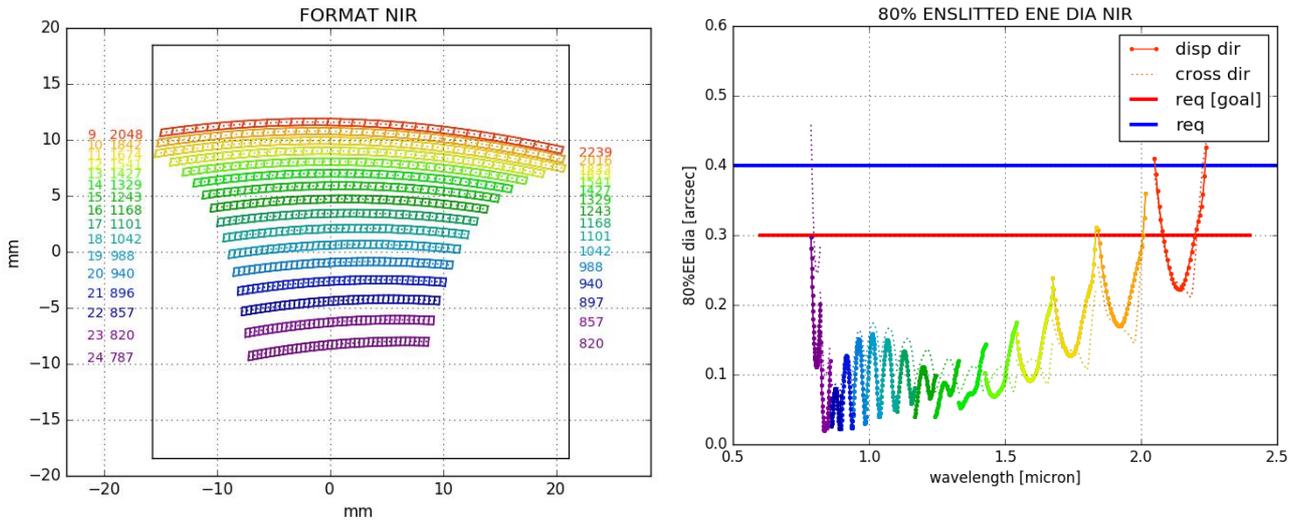

Figure 6. Left, spectral format. Right, enslitted energy diameter.

Example spot diagrams are presented in Figure 7 for different wavelengths in spectral order 15 for four different positions in the slit. Each spot is contained in a square of size 2 pixels or 36µm (0.5 arcsec).
The NIR detector is a HAWAII 2RG with a 2Kx2K, 18µm pixels and it will be operated at 40K in order to avoid possible persistence images.

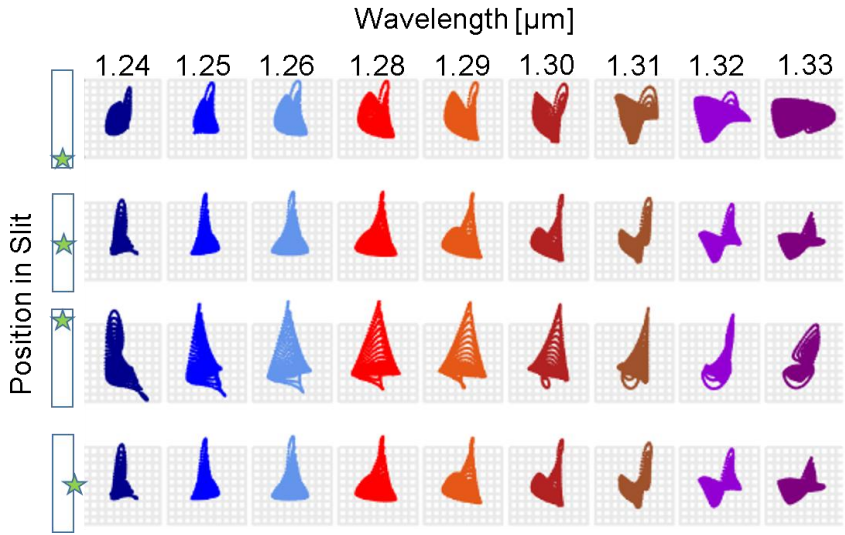

Figure 7. NIR spot diagrams for different positions in the slit. Order 15, 1244-1329 nm.

Ghosts images have been modeled in various ways. For the camera, which is axisymmetric and close to the detector, a sequential ghost analysis was applied. For the collimator and prisms, a non-sequential analysis was necessary. We do not expect major problems as the ratio of all the ghosts signal to the observations is about $10^{-4}$.

A summary of the main characteristics of the UV-VIS and NIR spectrographs is presented in Table 2.

Table 2. Spectrographs main characteristics.

|  | **UV-VIS** | **NIR** |
|---|---|---|
| Resolution Average (1" slit) | ~4700 | ~5300 |
| Slit Length | 12 arcsec | 12 arcsec |
| Spectral Range | 350-850 nm | 800-2000 nm |
| Detector | CCD 4k x 2k, 15µm pixel | H2RG 2k x 2k, 18 µm pixel |
| Working Temperature | Ambient | 150K, 40K detector |
| Throughput (sky-detector) | ~31% | ~26% |

## 5. CALIBRATION UNIT

The CU produces uniform light across the NIR and UV-VIS slits for flux and wavelength calibration. The light sources are: quartz-tungsten-halogen (QTH), deuterium (D2), ThAr and NeArHgXe penray lamps. The light is emitted out of an integrating sphere exit port, with the QTH and Penray lamps directly attached to the sphere while the D2 and ThAr lamps are injected through the remaining port of the sphere[2].

In order to replicate a uniform beam at the focal plane, with F/11 and pupil position as that of the telescope, the integrating sphere diaphragm is placed at the effective focal plane of the relay lens (Figure 8). With F/11 = f/D, the diaphragm diameter D can be calculated for the chosen relay lens with effective focal length f.

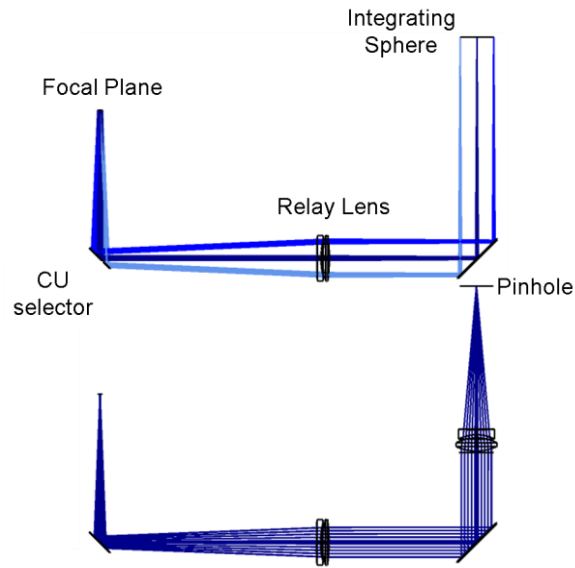

Figure 8. Calibration Unit optical layout. Top, illumination mode, bottom, synthetic star mode.

The resulting beam is uniform within the 1% level on a disk of diameter 4.8 mm at the focal plane and is larger than the equivalent 12-arcsec slit size of diameter 2.2 mm. The oversized region relaxes positioning tolerances.

The AC offers a synthetic star mode with a pinhole next to the integrating sphere and lenses to re-image it at the telescope focal plane (see bottom of Figure 8).

# 6. ACQUISITION CAMERA

The acquisition camera will be used for acquiring and centering the target field. It has broadband filters ugrizY (as in LSST) and V (Johnson) and a detector capable of science grade photometry. Depending on the AC selector position (Figure 9), the AC can perform other functions such as secondary guiding, slit viewing and synthetic star simulation[4].

The original requirements for the optical design are good image quality across the wavelength range from 350 to 950 nm (u to Y band), adequate field of view of 2 to 4 arcmin with a pixel scale smaller than ~1 arcsec/pixel. This was accomplished with the 5-element camera and singlet presented in Figure 9. The collimator lens, diameter 60mm, re-images the telescope pupil onto the camera, which consists of two doublets and two singlets. The camera (28mm maximum diameter) provides color correction for the whole wavelength range and relays the telescope focus on the detector (1Kx1K, 13μm pixel) with an F/3.6. The first slab of BK-7 glass simulates the filter while the last slab of Silica is the entrance window for the CCD dewar. The nominal image quality as measured by the spot size radius is about 5μm across the field of view of 3.5'x3.5' (0.2 arcsec/pixel). When we consider positioning and manufacturing errors, the spot size can be up to 15μm for 90% of 100 Monte Carlo simulations and it is still adequate for our pixel size.

The AC was designed to work at the expected mean temperature of 10C. When possible temperature changes of +/-10C are modeled, the image quality is degraded but still falls within 2x2 pixels. Furthermore, the image quality can be completely recovered by moving the collimator lens.

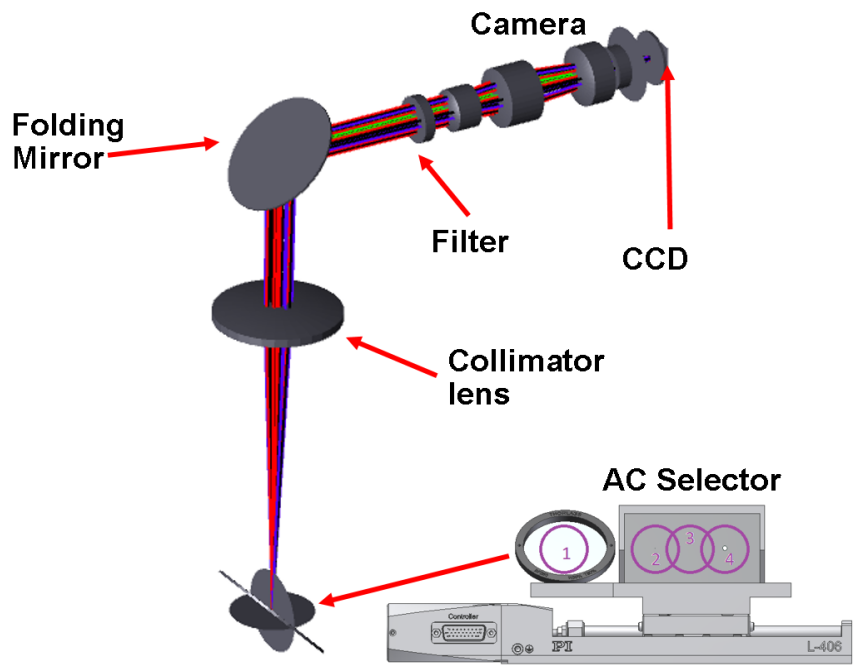

Figure 9. Acquisition Camera layout. The AC selector is illustrated (not to scale).

The overall throughput of the AC from sky to detector (detector excluded) is 27% assuming reflection coatings and AR efficiencies of 98% and elements-bulk-material absorption.

# 7. CONCLUSIONS

We have briefly described the optical design for the subsystems involved in the optics of SOXS: the Common Path, the NIR and UV-VIS spectrographs, the Calibration Unit and the Acquisition Camera. SOXS will be an instrument capable of observing simultaneously from 350 nm to 2000 nm with two spectrographs of medium resolution (R>4500). It will start operations in 2021 at the NTT telescope in la Silla.